\documentclass{nle}

\title[IATos]
      {IATos: study of an AI-powered pre-screening tool for COVID-19 from cough audio samples}
\author[D. Trejo Pizzo and S. Esteban]
       {David Trejo Pizzo\\
        Secretaría de Innovavión y Transformación Digital,\\
        Ciudad Autonoma de Buenos Aires, Argentina.
        \and
        Santiago Esteban\\
        Subsecretaría de Planificación Sanitaria, Ministerio de Salud,\\
        Ciudad Autonoma de Buenos Aires, Argentina.}

\received{7 December 2021}

\pagerange{\pageref{firstpage}--\pageref{lastpage}}
\pubyear{1998}

\begin{document}

\label{firstpage}
\maketitle

\begin{abstract}

\textbf{OBJECTIVE} Our objective is to evaluate the possibility of using cough audio recordings (spontaneous or simulated) to detect sound patterns in people who are diagnosed with COVID-19. The research question that led to our work was: what is the sensitivity and specificity of a machine learning based COVID-19 cough classifier, using RT-PCR tests as the gold standard?\\

\textbf{DESIGN} Diagnostic test validation study.\\

\textbf{SETTING} The audio samples that were collected for this study belong to individuals who were swabbed in the City of Buenos Aires in all public testing facilities and 1 private facility where RT-PCR studies were carried out on patients suspected of COVID, and 14 out-of-hospital isolation units for patients with confirmed COVID mild cases. The audios were collected through  the Buenos Aires city government WhatsApp chatbot that was specifically designed to address citizen inquiries related to the coronavirus pandemic (COVID-19).\\

\textbf{PARTICIPANTS} The first dataset collected corresponds to 2821 individuals who were swabbed in the City of Buenos Aires, between August 11 and December 2, 2020. Individuals were divided into 1409 that tested positive for COVID-19 and 1412 that tested negative. From this sample group, 52.6\% of the individuals were female and 47.4\% were male. 2.5\% were between the age of 0 and 20 , 61.1\% between the age of 21 and 40 , 30.3\% between the age of 41 and 60 and 6.1\% were over 61 years of age.\\

UPDATE: A second dataset containing 140.530 audio coughs was collected during the months of April to October 2021, and has 18.271 audios from individuals that tested positive for COVID-19 and 122.259 from individuals that tested negative.\\

\textbf{MAIN OUTCOME MEASURES} The audio files corresponding to each individual were categorized as "positive" and "negative" based on their PCR swab test results.\\

\textbf{RESULTS} Using the dataset of 2821 individuals our results showed that the neural network classifier was able to discriminate between the COVID-19 positive and the healthy coughs with an accuracy of 88\%. This accuracy obtained during the training process was later tested and confirmed with a second dataset corresponding to 492 individuals.\\

UPDATE: using a combination of the first and the second dataset, the first neural network used was re-trained. This new training involved using audio files from 143.351 individuals (2.821 from the first dataset and 140.530 from the second dataset). The neural network classifier was able to discriminate between the COVID-19 positive and the healthy coughs with an accuracy of 86\%. The performance also shows that this new trained NN is better at assessing negative coughs, and this may be due to the new unbalanced data that has 10\% of positive audio files and 90\% of negative audio files.\\

\textbf{CONCLUSION} Based on the reported results, we consider that the cough audio analysis could be used as a COVID-19 digital pre-entry screening tool to be used prior to conventional testing tools such as RT-PCR tests. This could help reduce wait times, ease stress, and lower exposure risk of healthcare personnel and other citizens.\\

\textbf{FUNDING STATEMENT} The project was carried out with the resources of the Secretariat of Innovation and Digital Transformation and the Ministry of Health of Buenos Aires City Government and with volunteers from other government areas for the field duties.\\

\textbf{AUTHOR DECLARATIONS}

All relevant ethical guidelines have been followed; any necessary IRB and or ethics committee approvals have been obtained and details of the IRB oversight body are included in the manuscript. \textbf{Yes}\\

All necessary patient - participant consent has been obtained and the appropriate institutional forms have been archived. \textbf{Yes}\\

The authors have no conflicts of interest to declare. All co-authors have seen and agree with the contents of the manuscript and there is no financial interest to report.
\end{abstract}

\section{Introduction}

As from November 28, 2021, the SARS-CoV-2 Coronavirus has spread to 213 countries around the world, has over 262 million confirmed cases and has caused over 5.2 million deaths \cite{1} \cite{22}.\\

The SARS-CoV-2 Coronavirus causes a respiratory infection called COVID-19.  The COVID-19 virus spreads primarily through droplets of saliva or discharge from the nose when an infected person coughs or sneezes \cite{36}.\\

Given that access to vaccines is limited and treatments are still in development, minimizing the spread through timely testing and isolating infected people is one of the best effective defenses humanity has at its disposal against COVID-19. However, the ability to deploy this strategy depends on each  country’s capability to test significant fractions of their populations, including those that are not yet in contact with the health system. Agile, scalable and proactive testing have become the key differentiator for managing this pandemic.

\subsection{Diagnostic strategies}

The main types of tests used for COVID-19 are:
\begin{itemize}
    \item Rapid serology antibody tests (ICT, LFA)
    \item Enzyme Linked Immunosorbent Assays (ELISA)
    \item Real-time polymerase chain reaction (RT-PCR) through swab tests
\end{itemize}

Rapid serology tests and ELISAs detect the presence of IgG / IgM antibodies produced by the immune system in response to a SARS-CoV-2 infection, while RT-PCR tests are molecular tests that detect the presence of viral RNA of SARS-CoV-2.\\

Triage remains as the fastest method for pre-screening a person, and this method is the one that can be scaled faster and at a low cost using exponential technologies. Triage is the process of sorting people based on their need for medical treatment as compared to their chance of benefiting from such care. Triage is usually performed in emergency rooms, disasters and wars, when limited medical resources must be allocated to maximize the number of survivors. During infectious disease outbreaks, triage is particularly important to separate patients likely to be infected with the pathogen of concern.\\

The main route of transmission of COVID-19 is through respiratory droplets generated when an infected person coughs or sneezes.  Any person who is in close contact with someone who has respiratory symptoms (e.g., sneezing, coughing, etc.) is at risk of being exposed to potentially infective respiratory droplets \cite{37}.\\

Recently, the US FDA \cite{18} has written about the importance of expanding rapid COVID-19 serological antibody testing. Current laboratory diagnostic tests for COVID-19 are based on labor-intensive molecular techniques (RT-PCR swab tests) and have generally been reserved for patients whose disease severity, age and/or comorbidities place them in high risk of serious illness, specially in countries with limited testing resources.\\

Recent work has also described that SARS-CoV-2 viral shedding often occurs in early pre-symptomatic stages, which could affect the dynamics and diagnostic accuracy of SARS-CoV-2 RT-PCR swab tests, especially if patients do not undergo molecular diagnostic tests until several days after the onset of symptoms. On the other hand, COVID-19 rapid serology antibody tests detect IgM/IgG antibodies, which are generated 1-3 weeks after the onset of symptoms.\\

Together, the SARS-CoV-2 RT-PCR swab test, combined with a rapid COVID-19 serological antibody test provides a more complete picture of COVID-19 disease progression, in a given individual and across populations.

\subsection{Importance of early diagnosis}

Many authors have highlighted that the key to safely returning back to normal life \cite{35}, especially while we wait for the COVID-19 vaccination process to be completed, does not lie in a single "perfect" COVID-19 test - but through multiple rapid and affordable tests and non-clinic pre-screening tools.\\

RT-PCR tests have been the gold standard for COVID-19 diagnosis. Nevertheless, their high cost and relatively long processing times make them a less appealing option  when deploying population wide testing strategies. On the other hand, antibody and antigen detection tests have proved to be an excellent complement to RT-PCR tests. Their lower costs, shorter processing times and higher availability allows for a more wide spread testing strategy and also shorter testing-isolation times.

\subsection{Difficulties with early diagnosis}

Currently, existing RT-PCR tests take considerably long to deliver results, are expensive to run frequently, or are often considered uncomfortable, which discourages people from going to get tested.\\

There are already some rapid tests for COVID-19 on the market. The Sofia SARS antigen test performed by the Quidel Corporation is approximately 99\% accurate, takes about 15 minutes to deliver results, and is already in use. A pharmacy in Arkansas, USA, reported that they had dispensed more than 800 in about two weeks. The test costs U\$D 95, which is a considerable price for most people’s budget, especially if they have to take the test periodically.\\

OpenCovidScreen \cite{17}, a coalition of scientists and biotech leaders working to drive innovation in COVID-19 testing and, in turn, accelerate return to work and school, says that  a good COVID-19 testing solution needs four items: frequency (at least weekly), more test points, an affordable price (around U\$D 10-20), and an easy method of collection (e.g. saliva, quick test, cough). Currently, there are tests available on the market that address one or two of these considerations, but there is nothing that brings all four together.

\subsection{Audio signals for pre-screening and disease monitoring}

Audio signals generated by the human body (e.g. sighs, breathing, heart, digestion) have often been used by physicians and clinical researchers in the diagnosis and monitoring of diseases. However, until recently, such signals were generally collected by manual auscultation during scheduled visits to a medical facility. Medical research has begun to use digital technology to collect body sounds (e.g. digital stethoscopes) and perform automated data analysis \cite{1}, (e.g. for wheezing detection in asthma \cite{2} \cite{3}).\\

Researchers have also been testing the use of the human voice to aid in the early diagnosis of a variety of diseases. For example, in Parkinson's disease the patient’s speech is analyzed for variations in smoothness to determine if there exists a lack of coordination in the vocal muscles \cite{4} \cite{5}. Also vocal tone, rhythm, and volume have also been explored to diagnose  illnesses such as post-traumatic stress disorder \cite{7}, traumatic brain injury, and some psychiatric conditions \cite{8}.\\

The use of human-generated audio as a biomarker for various diseases offers enormous potential for early diagnosis, as well as for affordable solutions that could be widely applied if integrated into devices such as cell phones. This is even more powerful if such solutions could monitor people throughout their daily lives in a discreet way.\\

Clinical work has focused on the use of voice analysis for specific diseases: for example, in Parkinson's disease, microphone and laryngoscope equipment have been used to detect the softness of speech resulting from lack of coordination on the vocal muscles \cite{6} \cite{12}. Voice characteristics have also been used to diagnose bipolar disorder; and to correlate pitch, rhythm, and volume with signs of invisible conditions such as post-traumatic stress disorder \cite{7}, traumatic brain injury, and depression. Voice frequency has been linked to coronary artery disease (as a result of hardening of the arteries that can affect voice production) \cite{6}. Companies such as Israel-based Beyond Verbal and Mayo Clinic have indicated in press releases that they are testing these approaches.\\

More recently, the microphone in basic devices such as smartphones and wearables has been used for sound analysis. In \cite{12}, the microphone is used to understand the context of the user and this information is added to generate a view of the surrounding environment of the recorded sound.. In \cite{13} the authors analyze the sounds emitted while the user is sleeping, to identify episodes of sleep apnea. Similar work has also used sound to detect asthma and wheezing \cite{2} \cite{3}.

\subsection{IATos: proposal and background}

Machine learning methods have been designed to recognize and diagnose respiratory diseases from sounds \cite{1} and more specifically cough: Bales et al.\cite{14} used convolutional neural networks (CNN) to detect cough within ambient audio and diagnose three potential diseases (bronchitis, bronchiolitis and whooping cough) based on its unique audio characteristics.\\

With the advent of COVID-19, researchers have begun to explore whether cough sounds could be diagnostic \cite{15}. A COVID-19-related cough detection study is presented in \cite{15} using a cohort of 48 patients tested with COVID-19 versus other pathological coughs, in which a set of models is trained. In \cite{11} the voice recordings of COVID-19 patients are analyzed to automatically classify the health status of the patients in four aspects, namely, the severity of the disease, the quality of sleep, fatigue and anxiety. Quatieri et al. \cite{26} showed that changes in vocal patterns could be a potential biomarker for COVID-19.\\

This research project aims to explore the use of human cough sounds as pre-diagnostic markers for COVID-19. Within the framework of this research study, we worked on the data acquisition of audio samples from people who quarantined in out-of-hospital isolation units for COVID-19 mild cases and close contacts or who were admitted to hospitals.\\

This research project differs from other similar initiatives in two main aspects. In the first place, the acquisition of audio samples was done through WhatsApp, and not through web pages. This allowed for a more streamlined data collection from the  study participants and simplified the signing of informed consents. On the other hand, only people who underwent a PCR or antigen swab test participated in the study. Also, their results had to be available in the Buenos Aires City Health Ministry COVID-19 results database. This helped us classify the audio samples of each individual for the first 5k audio dataset, according to the result of their RT-PCR results, considered as the current "gold standard" for COVID-19 diagnosis. The second dataset, which we used to re-train the neural network, has 140k audio files.  This second dataset is unbalanced (10\% corresponds to COVID-19 positive individuals and 90\% corresponds to COVID-19 negative individuals) and includes both audio files from people that had undergone an RT-PCR and antigen tests.\\

The expected outcome is to be able to analyze the patient’s cough as an additional marker that can be incorporated into the medical triage strategy. We seek to improve the pre-screening process that is carried out today prior to undergoing diagnosis testing methods such as PCR or rapid antigen tests. Having a digital Triage, which can be performed from anywhere using WhatsApp, allows us to exponentially scale the detection strategy for symptomatic and asymptomatic cases.

\section{Methodology}

Based on the objective of studying cough as a way to pre screen individuals for COVID-19, the research question that guided our work was: what is the sensitivity and specificity of a machine learning based COVID-19 cough classifier, compared to the standard diagnosis test for COVID-19?\\

In order to answer this question and prior to the development of the neural network, we consider the following steps as our methodology of work. The steps of this process are:

\begin{itemize}
    \item Generate a data acquisition process
    \item Generate a dataset
    \begin{itemize}
        \item Define the selection process for study participants
        \item Contact people who agree to participate in the study
        \item Start the audio acquisition flow
\end{itemize}
    \item Establish a preprocessing strategy for the collected audios
\end{itemize}

\subsection{Data description}

Observational or experimental studies require a priori sample size calculation in order to achieve the necessary statistical power to reject the null hypothesis \cite{25}. In the field of pattern recognition though, there is no standardized priori determination method to evaluate the generated models performance based on sample size of the datasets used for training.\\

Review of previous literature from past research uses an empirical approach through learning curves and the use of surrogate models to provide an answer to the problem of sample size \cite{26}. However, these ex-post-facto solutions do not take into account the nature of the classificatory problem to be solved, the complexity of the model used, the use of feature augmentation techniques, or the use of pre-trained models with the transfer learning method.\\

As an approach to determining the necessary sample size, we revised the literature that refers to the use of pattern recognition models applied to the audio signal produced by coughing. We reviewed pre-existing literature and  studied the differences of different samples sizes and their related metrics (see Table \ref{sample-table}).\\

\begin{table}[h]
  \caption{Bibliographic search of works that refer to the use of pattern recognition models applied to the audio signal produced by the act of coughing.}
  \begin{minipage}{\textwidth}
    \begin{tabular}{lcrrrrr}
    \hline\hline
    Author (Year) & Samples & COVID + & Others\footnote{Another non-COVID-19 positive respiratory disease.}& Control & S/E & AUC\\
    \hline
    Imran (2020) & 1838 & 70 & 226 & 247 & .96/.95 & NR\\
    \noalign{\vspace {.5cm}}
    Brown (2020) & 491 & 141 & 52 & 298 & NR & 0.875\\
    \noalign{\vspace {.5cm}}
    Bales (2020) & 268 & - & 268 & - & .89/.948 & NR\\
    \hline\hline
    \end{tabular}
    \vspace{-2\baselineskip}
  \end{minipage}
  \label{sample-table}
\end{table}

In light of the studied literature we established that a sample size of 5000 cough audio sounds could enable the correct performance of the generated model.\\

The data acquisition process  was approved by the Ethics Committee of the Elizalde Hospital (Comité de Etica - Hospital Elizalde). In August 2020 we started the process of collecting cough audios through the Buenos Aires City Government chatbot with the aim of creating an Open Voice dataset for COVID-19 cough discrimination.\\

We only used samples with two conditions:
\begin{itemize}
    \item The person providing the cough had undergone a PCR test.
    \item The audio files were recorded within 3 days of that PCR test.
\end{itemize}

At the end of the data acquisition process, we had collected 6,000 audio files, from 2,821 individuals. This collection of audio files was split into two groups:
\begin{itemize}
    \item Training and validation: 2,412 audio files from individuals that tested negative for COVID-19, and 2,477 audio files for individuals that tested positive These audio files belong to 2,329 individuals.
    \item Test: 995 audio files that belong to 492 individuals.
\end{itemize}

Each person included in the study had to generate a variety of representative audios of the act of coughing. We collected on average 3 coughs per subject accompanied by general subject information: age, sex, date of the swab/cough, location, outcome of medical diagnosis, and finally information regarding timing of onset of COVID-19 signs and symptoms (fever, tiredness, sore throat, difficulty breathing, persistent pain or pressure in the chest, diarrhoea and coughing).

\begin{center}
\begin{figure}[htp]
    \centering
    \includegraphics[width=12cm]{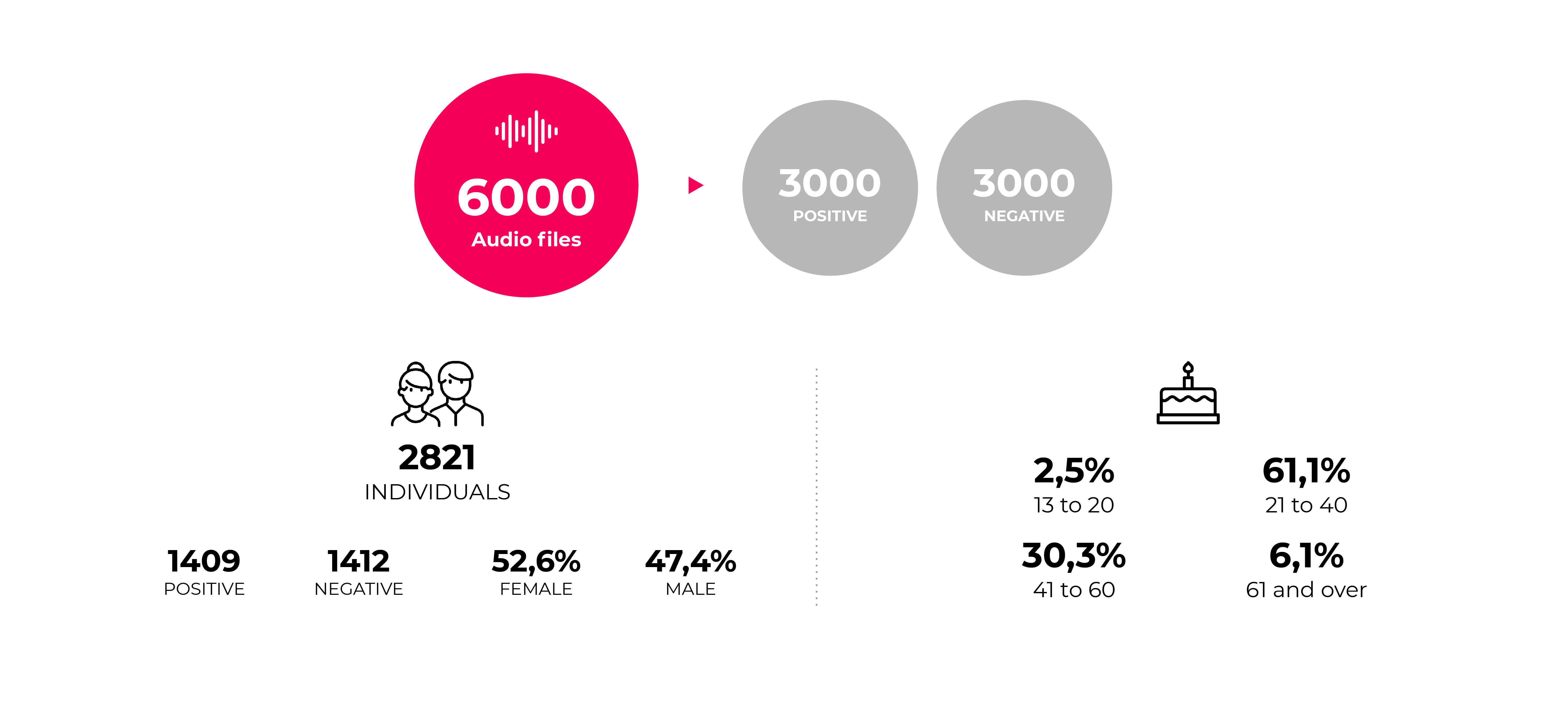}
    \caption{Dataset general information.}
    \label{fig:dataset}
\end{figure}
\end{center}

Data was anonymized before being collected on our secure server and samples were saved without compression in WAV format. Samples that had no coughs, too much noise or were silent were removed. No segmentation was performed on the cough recordings used to train and test the neural network.

\subsection{Data acquisition}

Individuals who participated in the study were asked to send one audio per day, for 3 consecutive days following the PCR test. The primary goal was to obtain multiple samples per person. Secondarily, we seek to study the evolution of audio patterns to detect if there are changes that show evidence of the evolution of the virus.\\

When someone went to a testing center, health system personnel invited them to participate in this study. If the person agreed, they were guided to start a conversational flow on the Buenos Aires city government WhatsApp chatbot specifically designed to address citizen inquiries related to the coronavirus pandemic (COVID-19). From the beginning of the study until September 30, 2020 people signed a printed informed consent to participate in this study. From September 30, 2020 the informed consent was digitized  into the WhatsApp chatbot conversation.

\begin{figure}[htp]
    \centering
    \includegraphics[width=12cm]{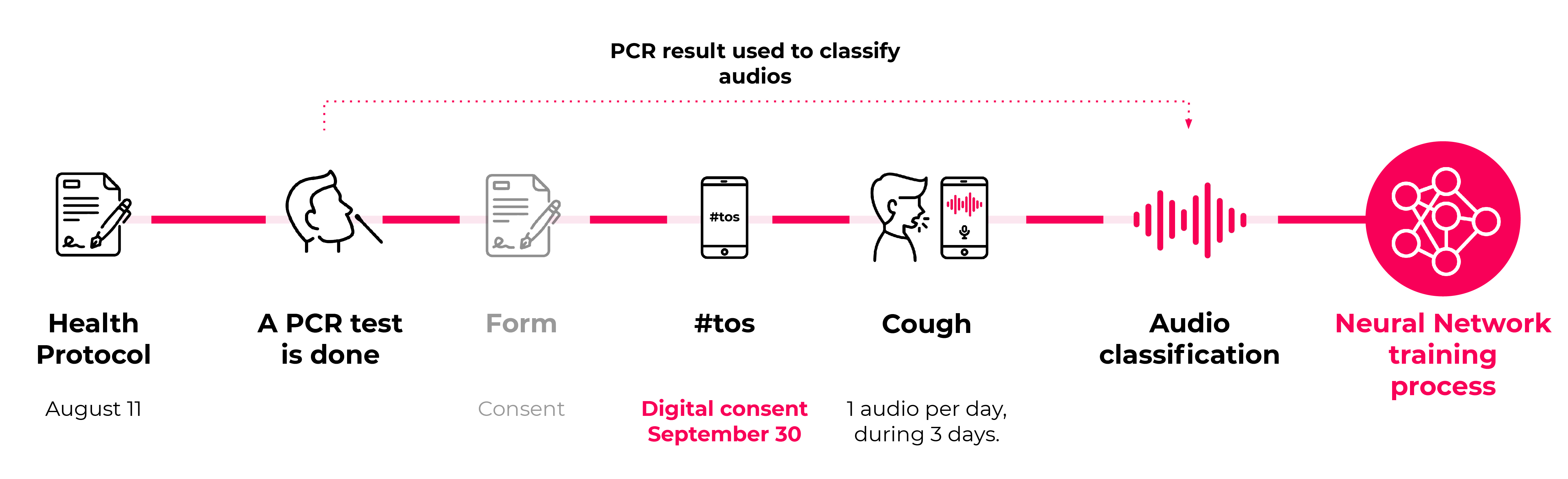}
    \caption{Data acquisition timeline.}
    \label{fig:timeline}
\end{figure}

The audio files that were collected for this study belong to individuals who were swabbed in the City of Buenos Aires in public and private facilities, out-of-hospital isolation units for patients with confirmed COVID mild cases and testing points of the strategic testing program "DetectAR".  The health personnel who performed the swab briefly explained the study to the individuals and invited them to participate.

\begin{figure}[htp]
    \centering
    \includegraphics[width=12cm]{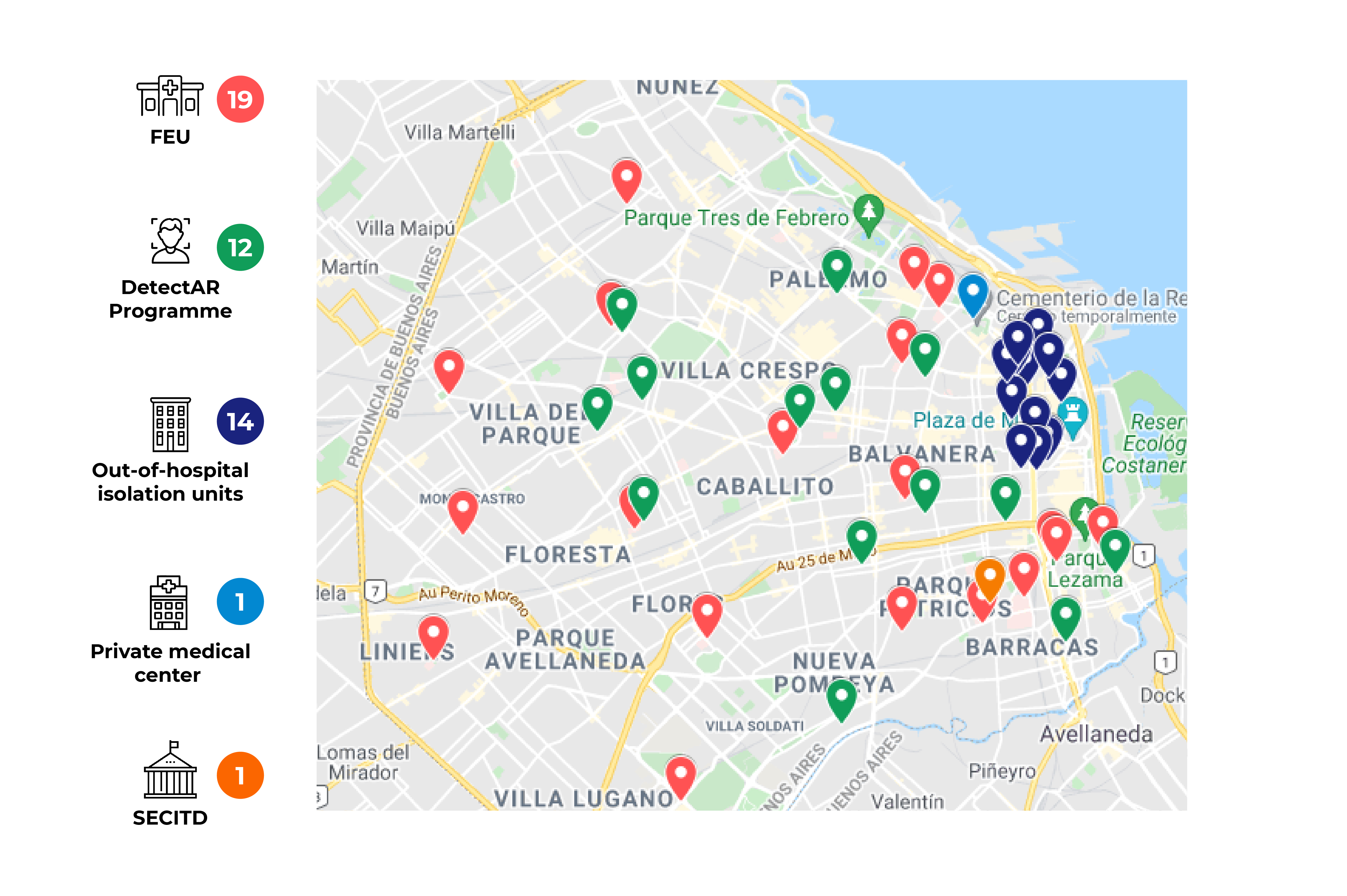}
    \caption{Data acquisition sites withing the Buenos Aires health system.}
    \label{fig:sites}
\end{figure}

Data collection sites:  19 Febrile Emergency Units (FEU), 12 testing points of the strategic testing program "DetectAR", 14 out-of-hospital isolation units for COVID-19 mild cases and close contacts and 1 private medical center as well as government personnel.

\subsubsection{Update on data acquisition}

A second dataset containing 140.530 audio coughs was collected during the months of April to October 2021, and has 18.271 audios from individuals that tested positive for COVID-19 and 122.259 from individuals that tested negative.\\

This second data acquisition process was carried on from April 2021 until October 2021, and the same method for audio collection was applied. In this case, we incorporated information related to the clinical triage.\\

Each individual answered if they had symptoms of Covid-19 at the time of sending the audio, if they were in close contact with someone with COVID-19, if they had breathing problems or if they were part of a risk group. Additionally, the data corresponding to the type of test that was carried out was stored as a variable for future studies. On this occasion, the samples were not analyzed looking for the percentage of men and women who participated, nor the age groups, assuming that the sample is a representation of the population of the City of Buenos Aires that uses the public testing system.

\subsection{Data preprocessing}

The preprocessing pipeline consists of sampling, denoising and cough discrimination from noise and silence audio chunks.\\

Although these could be extended to more stages, our current proposal (Fig. X) consists of:
\begin{itemize}
    \item Detects the presence of a cough in an audio, and cuts an audio segment of 1 second maximum duration.
    \item Normalize the audios in amplitude so that they all have a similar amplitude domain.
    \item Denoise audio files that have a low signal-to-noise ratio (SNR).
\end{itemize}

\begin{figure}[htp]
    \centering
    \includegraphics[width=12cm]{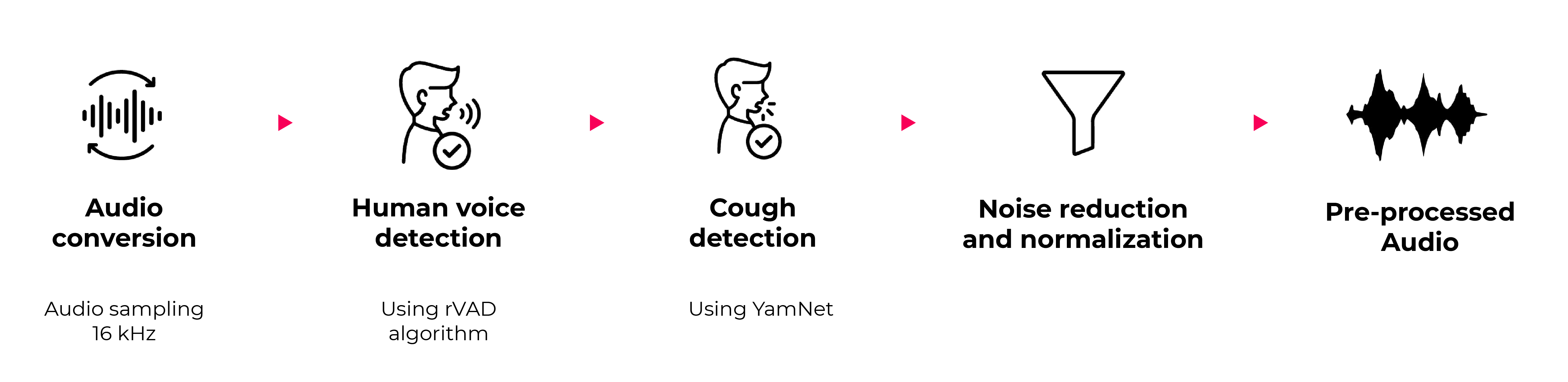}
    \caption{Pre-processing pipeline}
    \label{fig:prepro}
\end{figure}

As the first stage of preprocessing the audios, they are converted from the .ogg format to the .wav format, and all the audios are sampled equally at 16 kHz.\\

Subsequently, the audio files are cut into 1 second chunks. Then, using YAMNet and a network trained from the collected audio files, audio chunks that contain coughs are separated from those that are noise or silence. Finally, the resulting fragments are filtered to remove any noise that the signals may have.\\

The audios are then preprocessed in order to extract their main characteristics. This was achieved through a spectral analysis and selection of features of the signal. Finally the audio samples were normalized in amplitude.\\

From the information extracted, images were generated and introduced into a deep learning model, in order to be able to classify new cough audio signals.\\

For data processing, we mainly used a transformation that has broad consensus in the scientific community. In the first place, from the audio signal, the Mel \cite{27} spectrogram was obtained, extracting the most relevant characteristics of the sound using human auditory perception as a criterion. The transformation consists of splitting the signal into short frames, applying the discrete Fourier transform (DFT) to generate the spectrogram, then applying the Mel Filterbank -to calculate the energy in each Mel filter- and finally calculating the logarithm of the energy for each filterbank \cite{28}.\\

We used regularization techniques in our data since, given the complexity of the model to be applied, this will allow a better generalization of them. Specifically, to feed the model we used data augmentation, that is, we performed small random transformations to create new images, for example scaling and/or rotating the data. This technique allowed us to generalize the information and also obtain an improvement in the performance over audio signal data.

\section{Neural network model}

A deep learning-based COVID-19-associated cough detector was created. That is, a three-class classifier based on Convolutional Neural Networks that uses Mel's spectrograms as input information. Due to the subtle differences that exist between images of coughs corresponding to healthy patients, with various respiratory diseases, or with COVID-19, a neural network with more layers was required. Therefore, we used a CNN architecture.\\

\begin{figure}[htp]
    \centering
    \includegraphics[width=10cm]{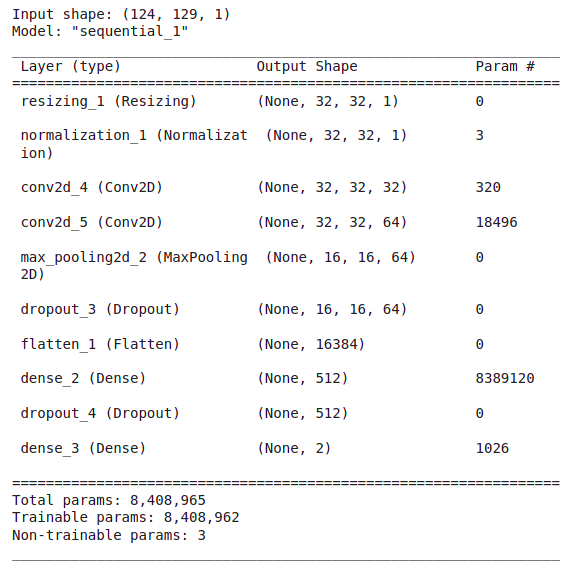}
    \caption{CNN Architecture.}
    \label{nn}
\end{figure}

Regarding the architecture, CNNs have a set of convolutional layers, to which pooling layers are then added and the dimensions of the matrix are adjusted. This is repeated successively and in the end, a fully connected layer is added to perform the prediction through a softmax.\\

For this project, three widely accepted scientific metrics will be used: Specificity, Recall/Sensitivity and Accuracy. In the training process of the model, a constant monitoring of the precision metric was carried out in order to generate the least amount of false positives. Due to the way COVID-19 spreads, it will always be a priority to maintain a high criterion of reliability and certainty before recommending a patient to seek assistance at a health facility. In parallel, the AUC measurement was monitored, which represents the area under the ROC curve that allows visualizing the sensitivity versus specificity.\\

We started by using a classification threshold of 0.5. Our proposed architecture, drawn in Fig. \ref{nn}, takes a recording with one or more coughs, performs pre-processing steps and inputs it into a CNN based model to output a pre-screening result. As pre-processing, each input audio file is split into 1 second audio chunks, padded as needed, processed with the MFCC package and subsequently passed through the neural network classifier. The output of these steps becomes the input to a CNN.\\

In order to train the classifier model, the set of spectrograms obtained were divided into train (80\%), validation (10\%) and internal test (10\%). Finally, with the model obtained, inferences were made to a second set of audio files that belong to a test set of 492 individuals. In this way we were able to evaluate the performance of the model on new data that it had never analyzed before.

\section{Results}

As we explained in the previous section, in order to evaluate the model we use the performance metrics of accuracy, specificity and sensitivity/recall on the test set. The accuracy here refers to the overall accuracy of the model. These performance metrics are based on mean confusion matrices from cross-validation. Tuning of the various hyper-parameters (number of hidden layers, learning rate, activation functions, dropout rate) of deep neural network-based models has also been performed, based on the cross-validation accuracy. Furthermore, the decay of model loss versus the number of epochs has been investigated to rule out the possibility of overfitting.

\subsection{Cough discrimination accuracy}

When we receive an audio file, the first step is to discriminate whether it has a cough or not. For this step we combine the use of YAMNet and a neural network that we specially trained to distinguish coughs. The analyzed audio file is cut into 1-second chunks and these fragments are evaluated by this detection stage, to discard the fragments that do not have the presence of cough. We achieved an accuracy of 97\% determining which fragments have a cough and which have noise compared to a random selected, manually classified sample of 100 1 second audio chunks. This allowed us to move on to the next stage, which is to determine if the audio fragments tagged as “cough” by the YAMNet algorithm,  correspond to that of an individual who is potentially COVID-19 positive or not.

\subsection{COVID-19 detection}

\begin{figure}[htp]
    \centering
    \includegraphics[width=10cm]{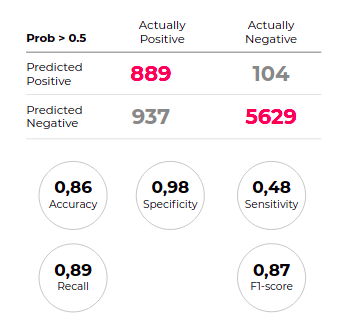}
    \caption{Confusion matrix.}
    \label{matrixnn}
\end{figure}

At the moment and based on the collected data, the overall accuracy of the deep learning based classifier is 86.00\%, Recall is 0.89 and F1-score is 0.87. The mean normalized confusion matrix resulting from this approach is shown in Fig. \ref{matrixnn}. Future work will continue to improve this model as more training data becomes available. Fig. XX shows the mean loss versus epochs of the neural network classifier, for both training and validation data sets. Both the curves start to saturate after around 70 epochs, indicating a reasonable learning time, without overfitting.

\section{Discussion and future work}

We have shown that COVID-19 can be discriminated with 86\% accuracy using only a forced-cough audio recording. We find most remarkable that our model performs better with coughs from positive patients than those from negative ones.\\

This first stage of developing the model focused on training it on a large dataset to learn good features for discriminating COVID-19 forced-coughs. The results when evaluating the testing set (made with coughs from individuals diagnosed with a validated PCR and antigen tests) serve as an indicator that the model would have similar accuracy when deployed. The next stage of this project is turning on the NN classifier together with the triage that is currently being used in the Buenos Aires City WhatsApp chatbot. We will also gather more quality data that can further train, fine-tune, and validate the model in a real world environment.\\

Since there might be cultural and age differences in coughs, future work could also focus on tailoring the model to different age groups and regions of the world using the metadata captured, and possibly including other sounds or input modalities such as vision or natural language symptom descriptions.

\subsection{Potential use cases of IATos}

Motivated by an urgent need, we developed a research protocol that was intended to evaluate the usefulness of an AI-based preliminary diagnostic tool for COVID-19, ubiquitously scalable through WhatsApp audio files containing a recording of an individual's forced cough.\\

The central idea of such a tool is inspired by previous studies \cite{10} \cite{16} showing that coughing can be used as a pre-screening tool for the diagnosis of a variety of respiratory diseases using AI techniques. The main objective is that the cough analysis is a complement and never a replacement for other diagnostic methods.\\

\begin{figure}[htp]
    \centering
    \includegraphics[width=10cm]{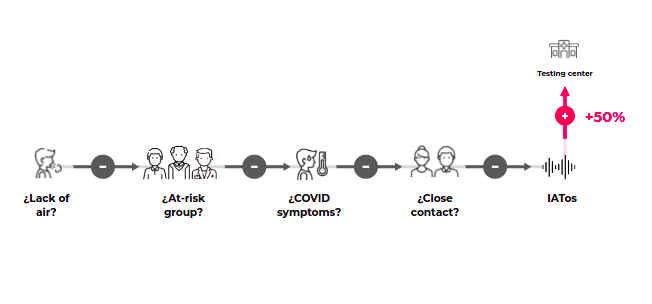}
    \caption{Triage flow with IATOS.}
    \label{triage}
\end{figure}

This non-invasive, free to access and real-time pre-screening tool may prove to have a great potential to complement current efforts to contain the disease spread in many different contexts.

Possible use cases could include:
\begin{itemize}
    \item Population daily screening tool: As workers go back to work, students go back to school, and commuters use public transport, to name a few, methods are required to screen infected COVID-19 carriers, especially asymptomatics. The only screening method currently available is using thermometers, however this study \cite{36} showed only 45\% of mild-moderate COVID-19 cases have fever (this represents 9\% of COVID-19 positives when asymptomatics are included). Meanwhile our AI tool  has shown to discriminate against a forced cough of COVID-19 positives with 86\% accuracy and could act as a complement to the medical triage questions.
    \item Pre-selection of candidates for test pooling: The test pooling strategy is expected to be employed in many countries, especially in low-incidence areas to rapidly identify a subgroup of individuals likely to be infected, however, “preliminary results show there is no dilution and no decrease on test sensitivity when mini pools of five samples each are used” \cite{37}. Group testing with our tool as shown in Fig. \ref{triage}, could pre-screen school classrooms, factories or even countries on a daily basis signalling probable infected candidate groups for smaller test pooling batches.
\end{itemize}

This technology could be assembled with low levels of effort and risk within the virtual triage that is carried out as a follow-up protocol for close contacts of positive patients.\\

In the future, subject to the learning curve and validation of the model, additional use cases could include \cite{10}:
\begin{itemize}
    \item Complementing temperature scanners at airports, borders or other key places where the virus circulates.
    \item Allowing remote pre-monitoring for anyone, anywhere, at any time, regardless of the existing infrastructure/testing facilities.
    \item Providing a centralized record of tests with location and time stamps. The data collected from the app could serve as input for long-term health care planning and health policy formulation.
\end{itemize}

Despite the good performance that is preliminarily observed in these tools, they are not intended as a replacement for clinical tests but as a complement. The goal is to investigate the potential for a single functional tool to monitor, track and control the rampant spread of the global pandemic in a timely, cost-effective and most importantly safe manner, by allowing a pre-test for anyone with access to a smartphone and an internet connection.

\section{Conclusion}

We presented an AI pre-screening tool that uses a neural network classifier that is able to discriminate COVID-19 positives with 86\%  accuracy from a forced-cough recording, at essentially no cost and with an exponential scaling capacity.\\

Despite its performance, IATOS is not meant to compete with clinical testing. Instead, it offers a unique functional tool for timely, cost-effective and most importantly safe monitoring, tracing, tracking and thus, controlling the rampant spread of the global pandemic by virtually enabling testing for everyone.\\

The proposed solution could be seen as a group outbreak detection tool for pre-screening whole-populations on a daily basis, taking the pre-screening to each house.. At the same time, this tool will allow us to send more people, potentially asymptomatic, to be tested. In this way we may be able to detect many positive asymptomatic cases that would normally go undetected.\\

As part of our ongoing pilots, data pipelines with FEUs and detection points have been set up to continue to improve the neural network. We plan on leveraging this data to further train and validate our models with the aim of improving pandemic management practices.\\

As part of our efforts to make data public and contribute to open innovation, we are opening our research methods and our dataset to collaborate and inspire others to develop similar tools to fight the COVID-19 pandemic, and hopefully be better prepared for the next pandemic.

\section{Ethical aspects}

The study will be guided by the rules of good clinical practice, the Declaration of Helsinki and the regulations of the Government of the City of Buenos Aires in force. Authorization was requested from the Research Ethics Committee, the DGDIYDP (Dirección General de Docencia, Investigación y Desarrollo Profesional) and the SSPLSAN (Subsecretaría de Planificación Sanitaria. Ministerio de Salud). Informed consent of those responsible, subjects and / or consent of minors will be requested as appropriate. Due to the context of the study and the specific SOPs of the Ethics Committee of the Elizalde Hospital (CEI-HGNPE, Hospital General de Niños Pedro de Elizalde), the documentation of the process will be protected in images of the forms duly signed by the subjects or their managers, who will keep the originals. The audio data of the patients and their COVID-19 test results will be stored without direct or indirect identifiers. Because it is an observational study and is framed in the context of the COVID-19 pandemic, the CEI-HGNPE requested its expeditious evaluation and approval.\\

\textbf{Data Availability} All the data referred to in the manuscript are available in https://data.buenosaires.gob.ar/.\\

\textbf{Declaration of competing interest} The authors declare that they have no known competing financial interests or personal relationships that could have appeared to influence the work reported in this paper.\\

\textbf{Acknowledgement} This work is dedicated to those affected by the COVID-19 pandemic and those who are helping to fight this battle in any way they can.\\

\appendix

\label{lastpage}
\end{document}